\documentclass[english,12pt]{article}
\usepackage[T1]{fontenc}
\usepackage[latin1]{inputenc}
\usepackage{geometry}
\usepackage{epsfig}
\usepackage{xspace}
\usepackage{cite}

\IfFileExists{url.sty}{\usepackage{url}}
                      {\newcommand{\url}{\texttt}}

\makeatletter

\usepackage{babel}
\makeatother
\date{}
\title{\huge \bf \textsc{\tt CutTools}:  a program implementing the {\tt OPP} 
reduction method to compute one-loop amplitudes
\\[1cm]}

\author{
{\bf 
 Giovanni Ossola$^{(1)}$ ,
 Costas G.~Papadopoulos$^{(1)}$ ,
 Roberto Pittau$^{(2,3)}$ 
}\\\\
{\normalsize $^{(1)}$Institute of Nuclear Physics, NCSR Demokritos,
15310 Athens, Greece}\\\\
{\normalsize $^{(2)}$Departamento de F\'{i}sica Te\'orica y del Cosmos}\\
{\normalsize Centro Andaluz de F\'{i}sica de Part\'{i}culas Elementales (CAFPE)}\\ 
{\normalsize Universidad de Granada, E-18071 Granada, Spain} \\\\
{\normalsize $^{(3)}$Dipartimento di Fisica Teorica}\\ 
{\normalsize Univ. di Torino and INFN sez. di Torino}\\
{\normalsize V. P. Giuria 1, I-10125 Torino, Italy}\\[1cm]
\texttt{http://www.ugr.es/{\small $\sim$}pittau/CutTools}
}
\begin{document}
\newcounter{im}
\setcounter{im}{0}
\newcommand{\exampleSp}{\stepcounter{im}\includegraphics[scale=0.9]{SpinorExamples_\arabic{im}.eps}}
\newcommand{\myindex}[1]{\label{com:#1}\index{{\tt #1} & pageref{com:#1}}}
\renewcommand{\topfraction}{1.0}
\renewcommand{\bottomfraction}{1.0}
\renewcommand{\textfraction}{0.0}
\newcommand{\nc}{\newcommand}
\nc{\eqn}[1]{Eq.~\ref{eq:#1}}
\nc{\be}{\begin{equation}}
\nc{\ee}{\end{equation}}
\nc{\ba}{\begin{array}}
\nc{\ea}{\end{array}}
\nc{\bea}{\begin{eqnarray}}
\nc{\eea}{\end{eqnarray}}
\nc{\bqa}{\begin{eqnarray}}
\nc{\eqa}{\end{eqnarray}}
\newcommand{\nl}{\nonumber \\}
\def\db#1{\bar D_{#1}}
\def\zb#1{\bar Z_{#1}}
\def\d#1{D_{#1}}
\def\tld#1{\tilde {#1}}
\def\slh#1{\rlap / {#1}}
\def\eqn#1{Eq.~(\ref{#1})}
\def\eqns#1#2{Eqs.~(\ref{#1}) and~(\ref{#2})}
\def\eqnss#1#2{Eqs.~(\ref{#1})-(\ref{#2})}
\def\fig#1{Fig.~{\ref{#1}}}
\def\figs#1#2{Figs.~\ref{#1} and~\ref{#2}}
\def\sec#1{Section~{\ref{#1}}}
\def\app#1{Appendix~\ref{#1}}
\def\tab#1{Table~\ref{#1}}

\maketitle
\begin{abstract}
We present a program that implements 
the {\tt OPP} reduction method to extract the coefficients of the one-loop 
scalar integrals from a user defined (sub)-amplitude or Feynman Diagram,
as well as the rational terms coming from the 4-dimensional 
part of the numerator.
The rational pieces coming from the $\epsilon$-dimensional part of the numerator
are treated  as an external input, and can be computed with the help
of dedicated tree-level like Feynman rules. 

Possible numerical instabilities are dealt with the help of 
arbitrary precision routines, that activate only when needed.
\end{abstract}

\newpage
\vspace*{\fill}
\tableofcontents
\vspace*{\fill}
\newpage

\section{Introduction}
Developing efficient tools to compute one-loop corrections for
multi-particle processes is an important task needed 
to cope with the complexity of LHC and ILC Physics.
In the last few years a big effort has been devoted by several authors to this
problem~\cite{Review}.
The used techniques range from analytic methods to
purely numeric ones, also including semi-numerical approaches. 

In the analytical approaches, computer algebra is used to reduce
generic one-loop integrals into a minimal set of scalar integrals
and remaining pieces (called rational terms), mainly by
tensor reduction~\cite{'tHooft:1978xw,Passarino:1978jh,Denner:2005nn,
Binoth:1999sp}. For
multi-particle processes this method becomes quite cumbersome
because of the large number of generated terms.

In the numerical or semi-numerical methods 
a direct computation of the tensor
integrals is performed~\cite{numerical}, capable, in
principle, to deal with any configuration of masses. However, their
applicability remains limited due to the high demand of
computational resources and the non-existence of an efficient
automation.

In a different approach, called the unitarity cut
method~\cite{unitarity-cut},
the one-loop amplitude rather than the
individual integrals are evaluated, avoiding the computation of 
Feynman diagrams. In another development,
the $4$-dimensional unitarity cut method has been used for the
calculation of QCD amplitudes~\cite{Britto:2004nc}, using
twistor-based approaches~\cite{twistors}. Moreover, a generalization
of the the unitarity cut method in $d$ dimensions, has been pursued
recently~\cite{Anastasiou}.
Nevertheless, in practice, only the part of the amplitude
proportional to the loop scalar functions can be obtained
straightforwardly. The remaining rational part, should
then be reconstructed either by using a direct computation based on
Feynman diagrams~\cite{Binoth1,Binoth2,Xiao} or by using a bootstrap
approach~\cite{Bern:2005cq}. Furthermore the complexity of the
calculation increases away from massless theories.

In two recent papers~\cite{Ossola:2006us,sixphotons}, 
we proposed a reduction
technique ({\tt OPP}) for arbitrary one-loop sub-amplitudes at {\it the
integrand level}~\cite{intlevel} 
by exploiting numerically the set of kinematical
equations for the integration momentum, that extend the quadruple,
triple and double cuts used in the unitarity-cut method. The method
requires a minimal information about the form of the one-loop
(sub-)amplitude and therefore it is well suited for a numerical
implementation. The method works for any set of internal and/or
external masses, so that one is able to study the full electroweak
model, without being limited to massless theories.
In~\cite{Ellis:2007br} the {\tt OPP} method has been used,
in the framework of the unitarity cut technique,
to explicitly compute the subtraction terms needed not to
double count the contribution of the various scalar integrals.

In this paper, we describe a {\tt FORTRAN90} implementation of
the {\tt OPP} algorithm. In section ~\ref{theory}, 
we recall the basics of the method and present our solution to
compute Rational Terms and to deal with numerical inaccuracies.
In section ~\ref{conventions} we outline the conventions
used in the program.
In section ~\ref{fortran} we describe the {\tt FORTRAN90} code that implements 
the method and, in the last section, we discuss our conclusions.
Finally, two appendices integrate the content of the paper.

\section{Theory and general features \label{theory}}
\subsection{The {\tt OPP} method}
The starting point of the {\tt OPP} reduction method is the general expression for the
{\it integrand} of a generic $m$-point 
one-loop (sub-)amplitude~\cite{Ossola:2006us}
\bqa
\label{eq:1}
A(\bar q)= \frac{N(q)}{\db{0}\db{1}\cdots \db{m-1}}\,,~~~
\db{i} = ({\bar q} + p_i)^2-m_i^2\,,~~~ p_0 \ne 0\,.
\eqa
In the previous equation, we use a bar to denote objects living
in $n=~4+\epsilon$  dimensions, and $\bar q^2= q^2+ \tld{q}^2$, where
$\tld{q}^2$ is $\epsilon$-dimensional and $(\tld{q} \cdot q) = 0$.
$N(q)$ is the $4$-dimensional part of the
numerator function of the amplitude. If needed,
the $\epsilon$-dimensional part of the numerator should be treated 
separately, as explained in~\cite{Pittausimple}.
$N(q)$ depends on the $4$-dimensional denominators
$\d{i} = ({q} + p_i)^2-m_i^2$ as follows
\bqa
\label{eq:2}
N(q) &=&
\sum_{i_0 < i_1 < i_2 < i_3}^{m-1}
\left[
          d( i_0 i_1 i_2 i_3 ) +
     \tld{d}(q;i_0 i_1 i_2 i_3)
\right]
\prod_{i \ne i_0, i_1, i_2, i_3}^{m-1} \d{i} \nl
     &+&
\sum_{i_0 < i_1 < i_2 }^{m-1}
\left[
          c( i_0 i_1 i_2) +
     \tld{c}(q;i_0 i_1 i_2)
\right]
\prod_{i \ne i_0, i_1, i_2}^{m-1} \d{i} \nl
     &+&
\sum_{i_0 < i_1 }^{m-1}
\left[
          b(i_0 i_1) +
     \tld{b}(q;i_0 i_1)
\right]
\prod_{i \ne i_0, i_1}^{m-1} \d{i} \nl
     &+&
\sum_{i_0}^{m-1}
\left[
          a(i_0) +
     \tld{a}(q;i_0)
\right]
\prod_{i \ne i_0}^{m-1} \d{i} \nl
     &+& \tld{P}(q)
\prod_{i}^{m-1} \d{i}\,. \eqa
Inserted back in \eqn{eq:1}, this expression
simply states the multi-pole nature of any $m$-point one-loop amplitude,
that, clearly, contains a pole for any
propagator in the loop, thus one has terms ranging from 1 to $m$ poles.
 Notice that the term with no poles, namely that one proportional to
$\tld{P}(q)$ is polynomial and vanishes upon integration
in dimensional regularization; therefore does not contribute to the amplitude,
as it should be.
  The coefficients of the poles can be further split in two pieces.
A piece that still depend on $q$ (the terms
$\tld{d},\tld{c},\tld{b},\tld{a}$), that vanishes upon integration,
and a piece that do not depend on q (the terms $d,c,b,a$).
 Such a separation is always possible, as shown in~\cite{Ossola:2006us}, and, with
this choice, the latter set of coefficients is therefore immediately
interpretable as the ensemble of the 
coefficients of all possible 4, 3, 2, 1-point
one-loop functions contributing to the amplitude.

 Once \eqn{eq:2} is established, the task of computing the one-loop amplitude
is then reduced to the algebraical problem of fitting
the coefficients $d,c,b,a$ by evaluating the function $N(q)$
a sufficient number of times, at different values of $q$,
and then inverting the system.
That can be achieved quite efficiently by singling out
particular choices of $q$ such that, systematically,
4, 3, 2 or 1 among all possible denominators $\d{i}$ vanishes.
 Then the system of equations is solved iteratively.
First one determines all possible 4-point functions,
then the 3-point functions and so on.
 For example, calling $q_0^\pm$ the 2 (in general complex) solutions for which
\bqa \d{0}= \d{1}= \d{2}=\d{3} = 0\,, \eqa (there are 2 solutions because
of the quadratic nature of the propagators) and since the functional
form of $\tld{d}(q;0123)$ is known, one directly finds the coefficient
of the box diagram containing the above 4 denominators through
the two simple equations
\bqa
N(q_0^\pm) &=& [d(0123) + \tld{d}(q_0^\pm;0123)] \prod_{i\ne 0,1,2,3}
\d{i} (q_0^\pm)
\,.
\eqa
This algorithm also works in the case of 
complex denominators, namely with complex masses.
 Notice that the described procedure can be performed
{\em at the amplitude level}. One does not need to
repeat the work for all Feynman diagrams, provided their sum is known:
we just suppose to be able to compute $N(q)$ numerically.

The modifications one has to apply to the method when 
working in $d = 4 + \epsilon$ dimensions are described in the next subsection.

 As a further remark notice that, since the terms
$\tld{d},\tld{c},\tld{b},\tld{a}$ still depend on $q$, also the
separation among terms in \eqn{eq:2} is somehow arbitrary.
 Terms containing a different numbers of denominators
can be shifted from one piece to the other in \eqn{eq:2},
by relaxing the requirement that the integral over
the terms containing $q$ vanishes. This fact provides an handle to cure
numerical instabilities occurring at exceptional phase-space points.
In {\tt CutTools} such a mechanism is implemented 
for the 2-point part of the amplitude, as described in 
subsection~\ref{numinacc}~. 

\subsection{The rational terms}
 The described procedure works in 4 dimensions.
However, even when starting from a perfectly finite tensor integral,
the tensor reduction may eventually lead to integrals
that need to be regularized \footnote{We use dimensional regularization
as a regulator.}.
 Such tensors are finite, but tensor reduction iteratively leads to
rank $m$ $m$-point tensors with $ 1 \le m \le 5 $, that are
ultraviolet divergent when $m \le 4$.
For this reason, we introduced, in \eqn{eq:1}, the $d$-dimensional
denominators $\db{i}$, that differs by an amount $\tld{q}^2$ from
their $4$-dimensional counterparts
\bqa
\db{i}= \d{i} + \tld{q}^2\,.
\eqa
The result of this is a mismatch in the cancellation
of the $d$-dimensional denominators of \eqn{eq:1} with the $4$-dimensional
ones of \eqn{eq:2}. The rational part of the amplitude, called 
$R_1$~\cite{paperrat}, comes from
such a lack of cancellation and is computed automatically in 
{\tt CutTools}.

A different source of Rational Terms, 
called $R_2$, can also be generated from the $\epsilon$-dimensional part 
of $N(q)$ (that is missing in \eqn{eq:1}), and should be added 
on the top of {\tt CutTools}'s results.
$R_2$ can be easily computed by using dedicated tree-level like Feynman rules, 
as explained in detail in~\cite{paperrat}. The user's conceptual effort
required to provide $R_2$ is the same needed to supply
the input function $N(q)$. We therefore consider the problem of computing $R_2$ 
completely trivial and solved once for all. 

The Rational Terms $R_1$ are generated by the following
extra integrals, introduced in~\cite{Ossola:2006us} 
\bqa \label{eq:ratexp} 
\int d^n \bar{q}
\frac{\tld{q}^2}{\db{i}\db{j}}             &=& - \frac{i \pi^2}{2}
\left[m_i^2+m_j^2-\frac{(p_i-p_j)^2}{3} \right]   +
\cal{O}(\epsilon)\,, \nl 
\int d^n \bar{q}
\frac{\tld{q}^2}{\db{i}\db{j}\db{k}}       &=& - \frac{i \pi^2}{2} +
\cal{O}(\epsilon)\,,\nl 
\int d^n \bar{q}
\frac{\tld{q}^4}{\db{i}\db{j}\db{k} \db{l}} &=& - \frac{i \pi^2}{6} +
\cal{O}(\epsilon)\,.\nl 
\eqa 
The coefficients of the above integrals are computed
in {\tt CutTools} by looking at the implicit mass
dependence (namely reconstructing the $\tld{q}^2$ dependence) in 
the coefficients $d,c,b$ of the one-loop functions, 
once $\tld{q}^2$ is reintroduced through the 
mass shift 
\bqa
\label{massshift}
m_i^2 \to m_i^2 -\tld{q}^2.
\eqa
One gets
\bqa 
b(ij;\tld{q}^2) &=&   b(ij)
                     +\tld{q}^2 b^{(2)}(ij) \,, \nl
c(ijk;\tld{q}^2) &=&   c(ijk)
                     +\tld{q}^2 c^{(2)}(ijk)\,.
\eqa 
Furthermore, by using \eqn{massshift}, the first line of \eqn{eq:2} 
becomes
\bqa
{\cal D}^{(m)}(q,\tld{q}^2) \equiv \sum_{i_0 < i_1 < i_2 < i_3}^{m-1}
\left[
          d( i_0 i_1 i_2 i_3 ;\tld{q}^2 ) +
     \tld{d}(q;i_0 i_1 i_2 i_3;\tld{q}^2)
\right]
\prod_{i \ne i_0, i_1, i_2, i_3}^{m-1} \db{i} \,,
\eqa
and the following expansion holds
\bqa
\label{bigd}
{\cal D}^{(m)}(q,\tld{q}^2)= \sum_{j= 2}^{m} \tld{q}^{(2j-4)} d^{(2j-4)}(q)\,,
\eqa
where the last coefficient is independent on $q$
\bqa
d^{(2m-4)}(q) = d^{(2m-4)}\,.
\eqa
In practice, once the $4$-dimensional 
coefficients have been determined, {\tt CutTools} redoes 
the fits for different values of $\tld{q}^2$, in order to determine 
$b^{(2)}(ij)$, $c^{(2)}(ijk)$ and  $d^{(2m-4)}$.
Such three quantities are the coefficients of the three
extra scalar integrals listed in \eqn{eq:ratexp}, respectively.

A different way of computing $ d^{(2m-4)}$ is implemented in {\tt CutTools}
when the {\tt Logical} variable {\tt inf} is set to {\tt .true.}
in {\tt subroutine dp\_get\_coefficients} and
{\tt subroutine mp\_get\_coefficients}.
In this case the code computes
\bqa
d^{(2m-4)} &=&  \lim_{\tld{q}^2 \to \infty}
\frac{{\cal D}^{(m)}(q,\tld{q}^2)}{\tld{q}^{(2m-4)}}\,.
\eqa
This limit is numerically quite stable and the computation faster.
However, the default for {\tt inf} is {\tt .false.}.
\subsection{Dealing with numerical inaccuracies \label{numinacc}}
During the fitting procedure to determine the coefficients, 
numerical inaccuracies may occur due to 
\begin{itemize}
\item[1)] appearance of Gram determinants in the solutions for which
          4, 3, 2 or 1 denominators vanish;
\item[2)] vanishing of some of the remaining denominators, 
          when computed at a given solution;
\item[3)] instabilities occurring when solving 
          systems of linear equations;
\end{itemize}
In principle, each of these three sources of instabilities
can be cured by performing a proper expansion 
around the problematic Phase-Space point \footnote{From now on we 
will denote such a point  as exceptional.}. 
An attempt in this direction is described in~\cite{sixphotons}.
However, this often results in a huge amount of work that, in addition, 
spoils the generality of the algorithm. Furthermore, one is anyway 
left with the problem of choosing a separation criterion
to identify the region where applying the proper expansion rather than
the general algorithm. 

 The solution implemented in {\tt CutTools} is, instead, of 
a purely numerical nature and
relies on a unique feature of the {\tt OPP} method: the fact that
the reduction is performed at the {\em integrand} level.
 In detail, the {\tt OPP} reduction is obtained when, as in \eqn{eq:2}, 
the numerator function $N(q)$ is rewritten in terms of denominators. 
Therefore $N(q)$ computed for some arbitrary value of $q$ by
using the l. h. s. of \eqn{eq:2} should always be {\em numerically} equal to
the result obtained by using the expansion in the r. h. s.
 This is a very stringent test that is applied in {\tt CutTools} 
for any Phase-Space point 
\footnote{The arbitrary, complex, 4-vector $q$ used for this test 
is randomly chosen by the code in a point by point basis.}. 
 When, in an exceptional Phase-Space point, these two numbers differ more
than a user defined quantity ({\tt limit}), 
the coefficients of the loop functions {\em for that particular point} 
are recomputed by 
using multi-precision routines (with up to 2000 digits)  
contained in {\tt CutTools}~\cite{multilib}.
 The only price to be payed by the user is writing, beside the
normal ones (namely written in double-precision), a multi-precision version
of the routines computing $N(q)$, that is anyway easily obtained
by just changing the definition of the variables used in the
routines, as explained in appendix ~\ref{appA}.
 The described procedure ensures that the coefficients of the scalar loop 
functions are computed with the precision given by {\tt limit}.
 This is usually sufficient; however, when strong cancellations
are expected among different loop functions, a multi-precision version of 
the one-loop scalar functions should also be used. Then, a complete
control over any kind of numerical inaccuracy is guaranteed.
Finally, one should mention that, usually, only very few points
are potentially dangerous, namely exceptional, so that a limited fraction of additional {\tt CPU} time
is used to cure the numerical instabilities, therefore compensating the fact
that the multi-precision routines are by far much slower than the normal ones.
This procedure has been shown to work rather well in practice.

A final remark is in order. For strictly 
massless momenta, all Phase-Space points are 
exceptional in the 2-point sector. Differently stated, expressing 
tensors such as
\bqa
\int d^n \bar{q} \frac{q^{\mu}}{\db{0}\db{1}}~~{\rm or}~~
\int d^n \bar{q} \frac{q^{\mu}q^{\nu}}{\db{0}\db{1}}
\eqa
in terms of scalar 2 and 1-point functions necessarily
involves the appearance  of powers of $\frac{1}{(p_1-p_0)^2}$, that is always
a problem when $(p_1-p_0)^2= 0$.

For this reason, a different basis~\cite{sixphotons} is implemented, 
for the 2-point sector, in {\tt CutTools}. 
This basis makes use of an arbitrary massless vector
$v$ and the code computes  the coefficients of the
following three scalar integrals
\bqa
\label{eq:newint}
\int\,d^n \bar{q} \frac{[(q+p_0)\cdot v]^\ell}{\db{0} \db{1}}
~~~~{\rm with}~~~\ell= 0,1,2~~{\rm and}~~v^2= 0\,.
\eqa
Notice that, when $k_1^2 \equiv (p_1-p_0)^2= 0$ and $m_0 = m_1 $, 
\bqa
\int\,d^n \bar{q} \frac{[(q+p_0)\cdot v]}{\db{0} \db{1}} &=&
- \frac{(k_1 \cdot v)}{2} \int\,d^n \bar{q} \frac{1}{\db{0} \db{1}}\,, \nl
\int\,d^n \bar{q} \frac{[(q+p_0)\cdot v]^2}{\db{0} \db{1}} &=& 
\frac{(k_1 \cdot v)^2}{3} \int\,d^n \bar{q} \frac{1}{\db{0} \db{1}}\,,
\eqa
exactly.
\section{Conventions used in the program \label{conventions}}
The information to be provided by the user is
\begin{itemize}
\item {\tt number\_propagators}   (integer)
\item {\tt rank}                  (integer)
\item {\tt num(q,qt2)}            (complex function)     
\item {\tt den0,den1,den2,den3,den4,den5} (derived types: see below)
\end{itemize}
The first variable refers to the number of propagators
in the (sub)-amplitude to be computed. 
The second variable is the maximum rank of $N(q)$ (not greater than\\
{\tt number\_propagators}, condition  that is anyway always fulfilled
in renormalizable gauges).
{\tt num(q,qt2)} is the numerator function $N(q)$, that, when pieces
of amplitude containing a different number of loop propagators are put 
together, also can depend on $\tld{q}^2$, that is the second entry
of the function {\tt num(q,qt2)}.

The last line of the above list refers to a derived type 
defined as follows
\begin{verbatim}
 module def_propagator
  implicit none
  type propagator
   integer :: i
   real(kind(1.d0)) :: m2
   real(kind(1.d0)), dimension(0:3) :: p
  end type propagator
 end module def_propagator      
\end{verbatim}
Therefore, ${\tt denj}$ contains the information
sufficient to denote the $j^{th}$
loop propagator, namely squared mass and 4-momentum. 
These loop propagators are internally classified 
according to a binary notation ${\tt denj} \to 2^j$ 
(following the user defined input ordering). 
The integer variable $i$ of the previous derived type, is internally set
to ${\rm i= 2^j}$ for each propagator.
In the present version of {\tt CutTools}, the maximum allowed number 
of loop propagators is six.
When less propagators are needed, they should be loaded
starting from the lowest value of ${\rm j}$.

At the end of the fitting procedure, the final results, namely
the coefficients of the scalar loop functions and the rational part
$R_1$, are loaded in the variables
\bqa
\label{eq:double1}
&&{\tt dcoeff(0,j)}\,,~
{\tt ccoeff(0,j)}\,,\nl
&&{\tt bcoeff(0,j)}\,,~
{\tt bcoeff(3,j)}\,,~
{\tt bcoeff(6,j)}\,~~{\rm and}~~{\tt rat1}.
\eqa
The second index labels the relevant scalar loop functions, according to the
above binary notation. For example
the coefficient of the 3-point function
\bqa
\int d^n \bar{q} \frac{1}{\db{0}\db{2}\db{4}}
\eqa  
is $\rm{ccoeff(0,2^0+2^2+2^4)} = \rm{ccoeff(0,21)}$ and that one of
\bqa
\int d^n \bar{q} \frac{1}{\db{1}\db{2}\db{3}\db{4}}
\eqa  
is $\tt{dcoeff(0,30)}$.
Furthermore, ${\tt bcoeff(0,j)}$,
${\tt bcoeff(3,j)}$, and ${\tt bcoeff(6,j)}$ are the coefficients of the
scalar integrals in \eqn{eq:newint} with $\ell= 0,1,2$, respectively.
When $\ell \ne 0$, also the knowledge of the vector $v$ is needed
\footnote{The massless vector $v$ is determined by {\tt CutTools}
in an event by event basis, to maximize the numerical stability.}. 
This information is stored in the array
\bqa
\label{eq:double2}
{\tt vvec(0:3,j)}\,,
\eqa
where the second index $j$ follows the same binary notation used for the
loop propagators.

Finally, when the multi-precision version of the code is activated,
the relevant output information is stored in the variables:
\bqa
\label{eq:multi}
&&{\tt mp\_dcoeff(0,j)}\,,~
{\tt mp\_ccoeff(0,j)}\,,\nl 
&&{\tt mp\_bcoeff(0,j)}\,,~
{\tt mp\_bcoeff(3,j)}\,,~
{\tt mp\_bcoeff(6,j)}\,,\nl
&&{\tt mp\_vvec(0:3,j)}\,~~{\rm and}~~{\tt mp\_rat1}.
\eqa

\section{Program structure \label{fortran}}
The directory structure looks as follows:
\begin{verbatim}
avh_olo_s4.f  dynamics.f90    MPREC        README
cuttools.f90  kinematics.f90  process.f90  tensors.f90
DOC           Makefile        rambo.f      type.f90
\end{verbatim}
In the following, we briefly discuss the content of each
file or directory in the previous list.
\subsection{avh\_olo\_s4.f}
This set of routines, provided by Andr\'e van Hameren, evaluates the 
scalar one-loop functions. In the current version
the fully massless scalar one-loop functions are included~\cite{andre}.
However, when needed, since {\tt CutTools} is not limited to massless processes,
a more general repository of one-loop master integrals can be 
used~\cite{repository}.  
\subsection{cuttools.f90}
It is the {\tt main} program. 
The distributed version implements, as a simple example, the reduction of
a five-point function with a ``toy'' numerator \footnote{The routines for
the evaluation of the complete one-loop QCD virtual corrections to the 
process $ q \bar q \to ZZZ$ will also be available from our webpage.}.
A test run output is given in appendix ~\ref{appB}.

The user should first 
initialize a few variables, such as the numbers of digits 
used by the multi-precision routines ({\tt idig}), filling
the internal tables of combinatorial factors by the calling
the subroutine {\tt load\_combinatorics}, setting the the 
number of propagators for the case at hand 
({\tt number\_propagators}), the maximum rank of $N(q)$ ({\tt rank}) and
the limit of precision below which the multi-precision routines activate
({\tt limit}).

Then, for each generated Phase-Space point  
(the maximum number of points {\tt nitermax}
should be provided at running time), the user should 
define the derived types {\tt denj} 
({\tt j = 0}, $\cdots$ , {\tt number\_propagators-1})
referring to the loop propagators, and load them by calling the
subroutine {\tt load\_denominators(den0,$\cdots$)}, 
with a number of arguments equal to the number of propagators.

Finally, the needed coefficients of the one-loop scalar functions
and the rational
part $R_1$ ({\tt rat1}) \footnote{We recall that the rational term 
$R_2$ ({\tt rat2}) should be computed separately.} 
are obtained by calling the subroutine {\tt get\_coefficients}.
At this point, if the precision test described in 
subsection~\ref{numinacc}~gives a result less then {\tt limit}, the program
multiplies all coefficients by the proper loop functions (this is achieved
by calling {\tt dp\_result(dbl\_prec,cutpart)}), adds the rational parts
and stores the event. 
Otherwise the entire procedure is repeated by using multi-precision.
If the test fails even using multi-precision (that may happen if
{\tt idig} is too small), 
the event is discarded.

At the end, the code, prints out the result of the Monte Carlo 
Phase-Space integration in the form of real and imaginary parts
of the finite term ({\tt sigma(0)}) and of the coefficients of
the $1/\epsilon$ ({\tt sigma(1)}) 
and $1/\epsilon^2$ ({\tt sigma(2)}) poles.
A statistics is also provided of the percentage of points 
computed with multi-precision or discarded. 
\subsection{DOC}
It is a directory containing this paper and any other updated documentation.
\subsection{dynamics.f90}
It is the part of the code where the user has to insert the numerator
function $N(q)$, namely the complex function {\tt num(q,qt2)}.
\subsection{kinematics.f90}
It is the core of {\tt CutTools}. It contains all routines 
needed to perform the fits. 
All the output variables listed in \eqn{eq:double1}, 
\eqn{eq:double2}  and \eqn{eq:multi} are located 
in {\tt module~coefficients}.
\subsection{Makefile}
 It is the {\tt Makefile} of {\tt CutTools}. The user should specify, among
other things, the {\tt FORTRAN90} compiler and the compilation flags 
he/she is using.
Notice that the multi-precision library in {\tt MPREC} 
should be compiled first (see next subsection).
\subsection{MPREC}
It is a directory containing the multi-precision package
of~\cite{multilib}. More precisely, before compiling {\tt CutTools},
the user should go to {\tt /MPREC/mpfun90/f90} and give the command
{\tt make} to compile the multi-precision library.
\subsection{process.f90}
All routines  needed to compute $N(q)$ should be put in this file.
\subsection{rambo.f}
It contains the random number generator and the routines for Phase-Space
generation, histogramming and bookkeeping of the events.
\subsection{README}
It is a {\tt .txt} file with information on the current version
of {\tt CutTools}.
\subsection{tensors.f90}
It contains the routines needed to perform scalar products of 4-vectors.
\subsection{type.f90}
It contains the {\tt FORTRAN90} derived types used by {\tt CutTools}.
\section{Conclusion}
We have presented {\tt CutTools}, a program implementing 
the {\tt OPP} reduction method~\cite{Ossola:2006us} to extract 
the coefficients of the one-loop scalar integrals 
from a user defined numerator function (namely (sub)-amplitude 
or Feynman Diagram), as well as the rational terms 
of type $R_1$~\cite{paperrat}.
The remaining part of the rational terms, $R_2$, should be supplied by the user
and can be computed with extra Feynman rules, as described in~\cite{paperrat}. 
The possible occurring numerical instabilities are treated with the help of 
arbitrary precision routines~\cite{multilib}. The {\tt OPP} algorithm
allowed us to implement a trivial check in order to activate the 
time consuming arbitrary precision routines only when necessary.

\section*{Acknowledgments}
We thank Andr\'e van Hameren for providing us with his private code
{\tt OneLOop} to compute massless one-loop scalar integrals.

G.O. and R.P. acknowledge 
the financial support of the ToK Program ``ALGOTOOLS'' (MTKD-CD-2004-014319).

C.G.P.'s and R.P.'s research was partially supported by the RTN 
European Programme MRTN-CT-2006-035505 (HEPTOOLS, Tools and Precision 
Calculations for Physics Discoveries at Colliders). 

The research of R.P. was also supported in part by MIUR under contract.
2006020509\_004.

C.G.P. and R.P. thank the  Galileo Galilei Institute for Theoretical
Physics for the hospitality and the INFN for partial support during the
completion of this work.

\section*{Appendices}
\appendix
\section{Going from double to multi-precision \label{appA}}
All routines in {\tt CutTools} have been written both in a 
{\rm normal} form (namely in double-precision) and in a multi-precision 
form.
Once a routine is written in normal form, the multi-precision
version of it can be easily obtained through the following changes 
in the declarations statements~\cite{multilib}:
\bqa
&&{\tt real(kind(1.d0)) \to type(mp\_real)}   \nl
&&{\tt complex(kind(1.d0)) \to type(mp\_complex)}\,.
\eqa 
The same strategy should be applied by the user 
to provide the multi-precision version of the routines to compute $N(q)$.
Finally, an {\tt interface} statement can be used to call both
versions with the same name.
\section{Test run output \label{appB}}
With   {\tt nitermax= 1}, the final output of the program reads as follows: 
\begin{verbatim}

 Result of the integration: 
         
 real_sigma(0)=  -67151075.5213172       +-    0.00000000000000     
 imag_sigma(0)=  -28426491.5346667       +-    0.00000000000000     
     
 real_sigma(1)=   3822974.11389803       +-    0.00000000000000     
 imag_sigma(1)=   3694493.63813566       +-    0.00000000000000     
     
 real_sigma(2)=  1.925254707235981E-029  +-    0.00000000000000     
 imag_sigma(2)= -1.427701757691647E-028  +-    0.00000000000000     
         
   Statistics on the mp routines:
         
 percentage of mp        points=   0.00000000000000     
 percentage of discarded points=   0.00000000000000     
        
 digits used in mp routines (if called) =          57

\end{verbatim}


\begin{thebibliography}{00}
\bibitem{Review}
  R.~K.~Ellis, W.~T.~Giele and G.~Zanderighi,
  JHEP {\bf 0605} (2006) 027
  [arXiv:hep-ph/0602185]; \\
  R.~Britto, B.~Feng and P.~Mastrolia,
  Phys.\ Rev.\  D {\bf 73} (2006) 105004
  [arXiv:hep-ph/0602178]; \\
  C.~F.~Berger, Z.~Bern, L.~J.~Dixon, D.~Forde and D.~A.~Kosower,
  Phys.\ Rev.\  D {\bf 74} (2006) 036009
  [arXiv:hep-ph/0604195]; \\
  Z.~Bern, N.~E.~J.~Bjerrum-Bohr, D.~C.~Dunbar and H.~Ita,
  JHEP {\bf 0511} (2005) 027
  [arXiv:hep-ph/0507019]; \\
  J.~Bedford, A.~Brandhuber, B.~J.~Spence and G.~Travaglini,
  Nucl.\ Phys.\  B {\bf 712} (2005) 59
  [arXiv:hep-th/0412108]; \\
  G.~Belanger {\it et al.},
  Phys.\ Lett.\  B {\bf 576} (2003) 152
  [arXiv:hep-ph/0309010]; \\
  A.~Denner, S.~Dittmaier, M.~Roth and M.~M.~Weber,
  Nucl.\ Phys.\  B {\bf 680}, 85 (2004)
  [arXiv:hep-ph/0309274]; \\
  A.~Denner, S.~Dittmaier, M.~Roth and L.~H.~Wieders,
  Nucl.\ Phys.\  B {\bf 724} (2005) 247
  [arXiv:hep-ph/0505042] and
  Phys.\ Lett.\  B {\bf 612} (2005) 223
  [arXiv:hep-ph/0502063];\\
  K.~Kato {\it et al.},
  PoS {\bf HEP2005} (2006) 312; \\
  T.~Binoth, T.~Gehrmann, G.~Heinrich and P.~Mastrolia,
  arXiv:hep-ph/0703311; \\
  S.~Weinzierl,
  arXiv:0707.3342 [hep-ph]; \\
  D.~Maitre and P.~Mastrolia,
  arXiv:0710.5559 [hep-ph]; \\
 Z.~Nagy and D.~E.~Soper,
  Phys.\ Rev.\  D {\bf 74} (2006) 093006
  [arXiv:hep-ph/0610028].
%
%
%
%

\bibitem{'tHooft:1978xw}
G.~'t Hooft and M.~J.~G.~Veltman,
Nucl.\ Phys.\ B {\bf 153} (1979) 365.
%
\bibitem{Passarino:1978jh}
G.~Passarino and M.~J.~G.~Veltman,
Nucl.\ Phys.\ B {\bf 160} (1979) 151.
%
\bibitem{Denner:2005nn}
  A.~Denner and S.~Dittmaier,
  Nucl.\ Phys.\ B {\bf 734} (2006) 62
  [arXiv:hep-ph/0509141] and
  Nucl.\ Phys.\ Proc.\ Suppl.\  {\bf 157} (2006) 53
  [arXiv:hep-ph/0601085].
%
\bibitem{Binoth:1999sp}
T.~Binoth, J.~P.~Guillet and G.~Heinrich,
Nucl.\ Phys.\ B {\bf 572} (2000) 361 [arXiv:hep-ph/9911342];\\
G.~Devaraj and R.~G.~Stuart,
Nucl.\ Phys.\  B {\bf 519} (1998) 483
[arXiv:hep-ph/9704308].
%
\bibitem{numerical}
  A.~Ferroglia, M.~Passera, G.~Passarino and S.~Uccirati,
  Nucl.\ Phys.\ B {\bf 650} (2003) 162
  [arXiv:hep-ph/0209219];
\\
W.~T.~Giele and E.~W.~N.~Glover,
arXiv:hep-ph/0402152;
%
\\
D.~E.~Soper,
Phys.\ Rev.\ D {\bf 62} (2000) 014009
[arXiv:hep-ph/9910292] and
Phys.\ Rev.\ D {\bf 64} (2001) 034018 [arXiv:hep-ph/0103262];
%
\\
Z.~Nagy and D.~E.~Soper,
JHEP {\bf 0309} (2003) 055 [arXiv:hep-ph/0308127].
%
\bibitem{unitarity-cut}
  Z.~Bern, L.~J.~Dixon, D.~C.~Dunbar and D.~A.~Kosower,
  Nucl.\ Phys.\ B {\bf 435} (1995) 59
  [arXiv:hep-ph/9409265]; \\
  Z.~Bern, L.~J.~Dixon, D.~C.~Dunbar and D.~A.~Kosower,
  Nucl.\ Phys.\ B {\bf 425}, 217 (1994); \\
  Z.~Bern, L.~J.~Dixon and D.~A.~Kosower,
  Annals Phys.\  {\bf 322} (2007) 1587
  [arXiv:0704.2798 [hep-ph]].


%
%
%
%
%
%
\bibitem{Britto:2004nc}
  R.~Britto, F.~Cachazo and B.~Feng,
  Nucl.\ Phys.\ B {\bf 725}, 275 (2005). 
%
\bibitem{twistors}
  E.~Witten,
  Commun.\ Math.\ Phys.\  {\bf 252}, 189 (2004);\\
  F.~Cachazo, P.~Svrcek and E.~Witten,
  JHEP {\bf 0409} (2004) 006;\\
  A.~Brandhuber, B.~J.~Spence and G.~Travaglini,
  Nucl.\ Phys.\ B {\bf 706}, 150 (2005);\\
  F.~Cachazo, P.~Svrcek and E.~Witten,
  JHEP {\bf 0410}, 074 (2004);\\
  I.~Bena, Z.~Bern, D.~A.~Kosower and R.~Roiban,
  Phys.\ Rev.\ D {\bf 71}, 106010 (2005).
%
\bibitem{Anastasiou}
  C.~Anastasiou, R.~Britto, B.~Feng, Z.~Kunszt and P.~Mastrolia,
  JHEP {\bf 0703} (2007) 111
  [arXiv:hep-ph/0612277]; \\
  C.~Anastasiou, R.~Britto, B.~Feng, Z.~Kunszt and P.~Mastrolia,
  Phys.\ Lett.\  B {\bf 645} (2007) 213
  [arXiv:hep-ph/0609191]; \\
  D.~Forde,
  Phys.\ Rev.\  D {\bf 75} (2007) 125019
  [arXiv:0704.1835 [hep-ph]]; \\
  N.~E.~J.~Bjerrum-Bohr, D.~C.~Dunbar and W.~B.~Perkins,
  arXiv:0709.2086 [hep-ph].
%
\bibitem{Binoth1}
  T.~Binoth, J.~P.~Guillet, G.~Heinrich, E.~Pilon and C.~Schubert,
  JHEP {\bf 0510} (2005) 015
  [arXiv:hep-ph/0504267]. 

\bibitem{Binoth2}
  T.~Binoth, J.~P.~Guillet and G.~Heinrich,
  JHEP {\bf 0702} (2007) 013
  [arXiv:hep-ph/0609054].
%
\bibitem{Xiao}
  Z.~G.~Xiao, G.~Yang and C.~J.~Zhu,
 Nucl.\ Phys.\  B {\bf 758} (2006) 1
  [arXiv:hep-ph/0607015];
\\
X.~Su, Z.~G.~Xiao, G.~Yang and C.~J.~Zhu,
  Nucl.\ Phys.\  B {\bf 758} (2006) 35
  [arXiv:hep-ph/0607016].
%
\bibitem{Bern:2005cq}
  Z.~Bern, L.~J.~Dixon and D.~A.~Kosower,
  Phys.\ Rev.\  D {\bf 73} (2006) 065013 [arXiv:hep-ph/0507005]; \\
  S.~D.~Badger, E.~W.~N.~Glover and K.~Risager,
  JHEP {\bf 0707} (2007) 066
  [arXiv:0704.3914 [hep-ph]].
\bibitem{Ossola:2006us}
  G.~Ossola, C.~G.~Papadopoulos and R.~Pittau,
  Nucl.\ Phys.\  B {\bf 763} (2007) 147 [arXiv:hep-ph/0609007].

\bibitem{sixphotons}
  G.~Ossola, C.~G.~Papadopoulos and R.~Pittau,
  JHEP {\bf 0707} (2007) 085
  [arXiv:0704.1271 [hep-ph]].

\bibitem{intlevel}
  F.~del Aguila and R.~Pittau,
  JHEP {\bf 0407} (2004) 017
  [arXiv:hep-ph/0404120] and
  R.~Pittau,
  arXiv:hep-ph/0406105. 
\bibitem{Ellis:2007br}
  R.~K.~Ellis, W.~T.~Giele and Z.~Kunszt,
  arXiv:0708.2398 [hep-ph].
\bibitem{Pittausimple}
  R.~Pittau,
  Comput.\ Phys.\ Commun.\  {\bf 104}, 23 (1997)
  [arXiv:hep-ph/9607309] and {\bf 111} (1998) 48
  [arXiv:hep-ph/9712418].

\bibitem{paperrat}
  G.~Ossola, C.~G.~Papadopoulos and R.~Pittau,
  arXiv:0802.1876 [hep-ph].
\bibitem{multilib}
D. H. Bailey; ARPREC (C++/Fortran-90 arbitrary precision package) 
http://crd.lbl.gov/{\small $\sim$}dhbailey/mpdist/. \\
See also D. H. Bailey, "A Fortran-90 Based Multiprecision System," ACM Transactions on Mathematical Software, vol. 21, no. 4 (Dec 1995), pg. 379-387.

\bibitem{andre}
  A.~van Hameren, J.~Vollinga and S.~Weinzierl,
  Eur.\ Phys.\ J.\  C {\bf 41} (2005) 361 [arXiv:hep-ph/0502165].

\bibitem{repository}
http://qcdloop.fnal.gov/. See also 
  R.~K.~Ellis and G.~Zanderighi,
  arXiv:0712.1851 [hep-ph];
  J.~R.~Andersen, T.~Binoth, G.~Heinrich and J.~M.~Smillie,
  arXiv:0709.3513 [hep-ph].
\end{thebibliography}
\end{document}